\documentclass[12pt]{article}
\usepackage{graphicx} 
\usepackage[english]{babel}
\usepackage{amsmath}
\usepackage{anysize}
\usepackage{hyperref}
\usepackage{comment}
\usepackage{multirow}
\usepackage{appendix}
\usepackage[utf8]{inputenc}
\usepackage{t1enc}
\usepackage{setspace}
\usepackage{indentfirst}
\usepackage{amsmath, amsthm, amssymb}
\usepackage{epstopdf}
\usepackage{subfigure}
\usepackage{appendix}
\usepackage[bottom]{footmisc}
\usepackage{float}
\usepackage{cite}
\usepackage{underlin}
\usepackage{fancyhdr}
\usepackage{epigraph}
\usepackage{authblk}
\usepackage{url}
\marginsize{2cm}{2cm}{1.5cm}{1.5cm} 
\usepackage{tikz}
\usepackage{graphicx}
\usepackage{cite}

\title{Charged particle multiplicity distributions derived from the Principle of Maximal Entropy%
\thanks{Presented at XXXI Cracow Epiphany conference on the Recent LHC Results, Cracow, 13-17 January 2025}%
}
\author{ S\'andor L\"ok\"os$^{1,2}$}
\affil{\small\textit{$^1$Institute of Nuclear Physics
Polish Academy of Sciences, Krak\'ow PL-31-342, Poland}}
\affil{\small\textit{$^2$Hungarian University of Agriculture and Life Sciences, Institute of Technology, Gyöngyös H-3200, Hungary}}
\date{}
\begin{document}
\maketitle
\begin{abstract}
Recent theoretical results renewed the interest in charged particle multiplicity distributions. The Shannon entropy of such distributions is conjectured to be related to the entanglement or von Neumann entropy of partonic quantum system. In this paper, we show that the measured charged particle multiplicities can be derived from the principle of maximum entropy (POME or MAXENT) without any \textit{a priori} physical assumption. The approach provides a natural explanation for the well-known negative binomial shape of the measured distributions.
\end{abstract}

\section{Introduction}

It has been theorized that considering the proton as a (maximally) entangled partonic quantum system could explain many puzzling theoretical and experimental observations \cite{Kharzeev:2017qzs,Baker:2017wtt,Tu:2019ouv}. The proton's wave function is \textit{partially} probed by another proton or electron, leading to entanglement between the probed and unprobed regions. The proposed observable is the entanglement or von Neumann entropy, which quantifies the degree of ``entanglement'' within the quantum system.

A series of papers aimed to describe the phenomenon and provides evidence to quantum entanglement between the sampled and non-sampled parts of the proton in the data by utilizing partonic dipole picture of pQCD \cite{Tu:2019ouv,Hentschinski:2021aux,Kharzeev:2021yyf,Zhang:2021hra,Hentschinski:2022rsa,Hentschinski:2023izh}. The calculations were compared to multiplicity measurements from proton-proton \cite{CMS:2010qvf,ATLAS:2016qux,ALICE:2017pcy} and electron-proton \cite{H1:2020zpd} collisions, showing convincing agreement between the data and the theoretical predictions.

Inspired by the success of the assumption of maximal entropy of the initial partonic system, it is worth investigating the problem from the final state perspective. The ultimate goal of this paper is to describe the measured charged particle distributions using the general principle of maximal Shannon entropy (POME), which has proven to be a powerful tool for addressing classical statistical physics problems in classical \cite{PhysRev.106.620} and in quantum systems \cite{PhysRev.108.171}, too. he essential idea is that Shannon entropy serves as the primer quantity, subject to certain constraints dictated by the data itself. Then the entropy is maximized subject to the constraints and a distribution is obtained that can be regarded as the least bias choice and contains the least amount of \textit{a priori} assumptions about the physical system. Thus, the use of POME is not the application of physical laws, but a method of reasoning that ensures that no tacit arbitrary assumptions are introduced \cite{PhysRev.106.620}. A similar approach was adopted elsewhere, see e.g. Refs. \cite{Malaza:1991pj,Navarra:2003um} and references in them.

\section{Charged particle multiplicities from POME}

The multiplicity of charged particles is, by nature, a statistical observable by its nature. The distribution of multiplicities can be interpreted as probability a probability distribution and is therefore normalized to unity. This interpretation can be regarded a constraint. With no further constraints, an assumption of a previous work can be recovered \cite{Hentschinski:2021aux}. The final state distributions can be directly derived from POME introducing an extra constraint that fixes the mean of the distribution.

\subsection{Initial state parton distributions}

If the initial partonic system is maximally entangled, it can consists of $N$, equally probable states, i.e., the probability of the partonic configurations is uniformly distributed: $p(n)=\frac{1}{N}$. Substituting this probability into Shannon's entropy formula, the entropy of the system can be derived. The number of states, $N$, therefore the entropy $S_\textmd{parton}$, too, can be expressed with gluon ($g(x,Q)$) and quark ($\Sigma(x,Q)$) distribution function as
\begin{equation}
    S_\textmd{parton} = - \sum_n p(n) \ln p(n) = \ln(N) = \ln(xg(x,Q)+x\Sigma(x,Q)),
\label{eq:Sparton_max}
\end{equation}
where $x$ is the Bjorken-$x$ and $Q$ is the virtuality. Hence $S_\textmd{parton}$ can be obtained as described in Refs.
\cite{Tu:2019ouv,Hentschinski:2021aux,Kharzeev:2021yyf,Zhang:2021hra,Hentschinski:2022rsa,Hentschinski:2023izh}. Let us emphasise that the assumption of the maximal entanglement is expressed through the probability $p(n)=\frac{1}{N}$ which leads to the result in Eq. (\ref{eq:Sparton_max}).

Now, instead of assuming this distribution, let us derive it from the principle of maximum entropy (POME) using the variational method. The functional to be maximized is given by
\begin{equation}
    \mathcal{F}_0[p,\alpha]=-\int_0^\infty p(n)\ln(p(n)) dn + \alpha \left(\int_0^\infty p(n) dn - 1\right),
\label{eq:F0}
\end{equation}
where the first term is Shannon entropy and the second term, with $\alpha$ being a Lagrange multiplier, enforces the normalization condition, ensuring that $p(n)$ is a probability distribution. Since it is the least constrained scenario, we shall denote it with the subscript $_0$. By performing the variation and utilizing the constraint, the probability can be expressed as
\begin{equation}
    \frac{\delta\mathcal{F}}{\delta p} \Rightarrow p(n) = e^{\alpha-1} \hspace{0.7cm} , \hspace{0.7cm} e^{\alpha-1}\int_0^\infty dn = 1 \Rightarrow p(n) = \frac{1}{N}.
\end{equation}
Entropy is chosen as the fundamental concept, and by maximizing it, the probability distribution (the uniform distribution) of the initial-state partonic system can be derived. Now, let's see how the final state distributions can be obtained with an additional constraint.

\subsection{Final state hadron distributions}

The same calculation can be repeated to derive the final state multiplicity distributions, however, an additional constraint needs to be introduced. In the measurements, the multiplicities are observed to have a mean (denoted by $\mu$). Therefore, an additional constraint can be included using a Lagrange multiplier $\beta$, as follows
\begin{equation}
    \mathcal{F}[p,\alpha,\beta]{=}\mathcal{F}_0[\alpha,p]{+}\beta \left(\int_0^\infty n p(n) dn {-} \mu \right),
\end{equation}
where $\mathcal{F}_0[\alpha,p]$ is defined in Eq. (\ref{eq:F0}). Using the variational method, the probability can be calculated as previously
\begin{equation}
    p(n)=\lambda e^{-\lambda n} \Rightarrow S_\textmd{hadron}=1-\ln(\lambda)
\label{eq:exp_pn}
\end{equation}
where $\lambda=1/\mu$ is the rate parameter of the distribution. This result can be considered to be a prediction and compared to the data with a free parameter to be fitted. Let us point out that the mean is constant, which in reality, is a good approximation for measurements performed in narrow rapidity ranges.

LHCb measured \cite{LHCb:2014wmv} charged-particle multiplicities in disjoint, fixed-width rapidity windows of 0.5 units, selected within the range of $2<\eta<4.5$. This approach provides the multiplicity distributions, hence their entropy as a function of rapidity. ALICE, ATLAS, CMS (see e.g. in Refs \cite{CMS:2010qvf,ATLAS:2016qux,ALICE:2017pcy})  as well as earlier experiments such as, e.g., at UA5 \cite{UA5:1988gup} or E665 \cite{E665:1993trt}, measured multiplicities cumulatively over wider and wider rapidity intervals, typically centered at $\eta=0$.

\section{Comparison to data}

\begin{figure}[htb]
\centerline{%
\includegraphics[width=0.45\textwidth]{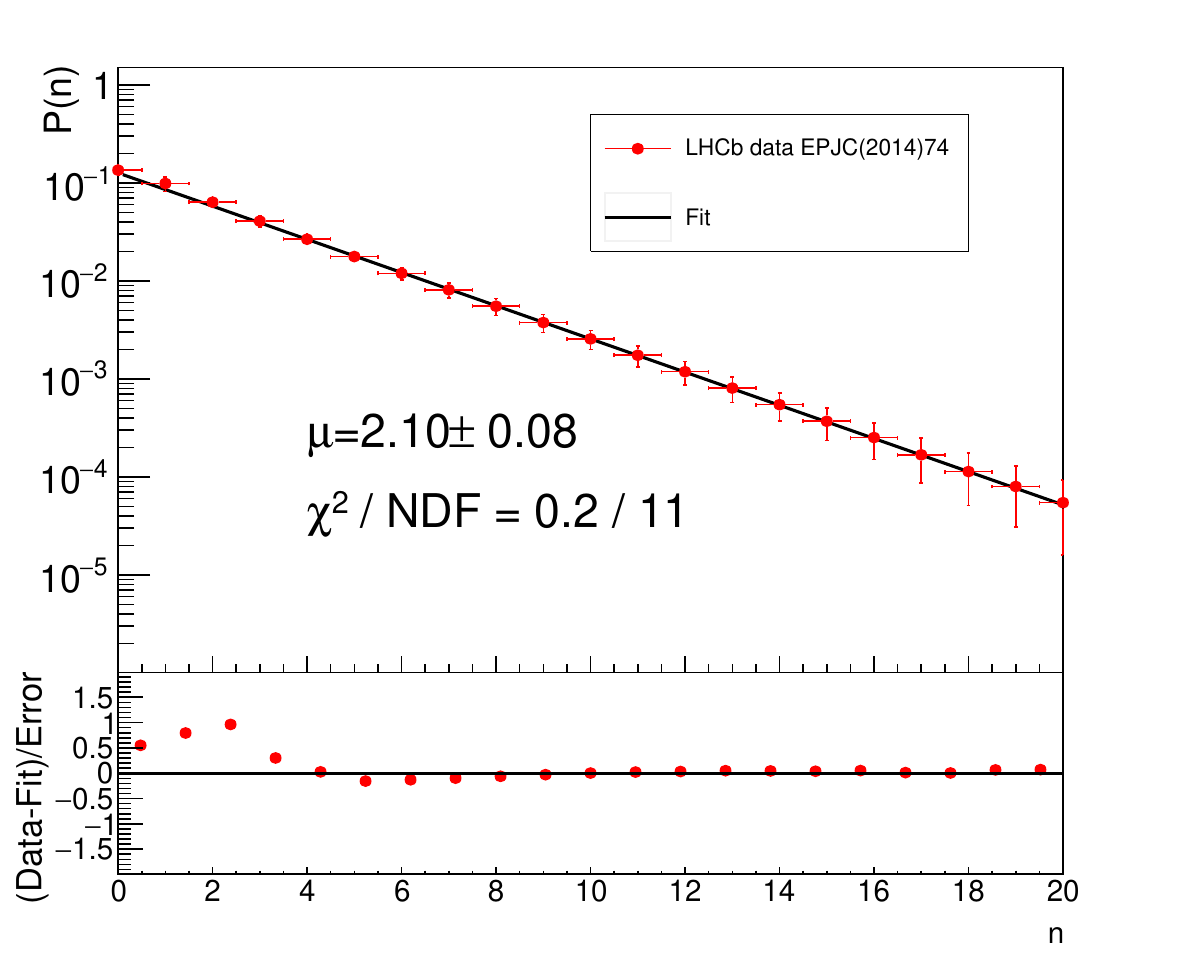}
\includegraphics[width=0.45\textwidth]{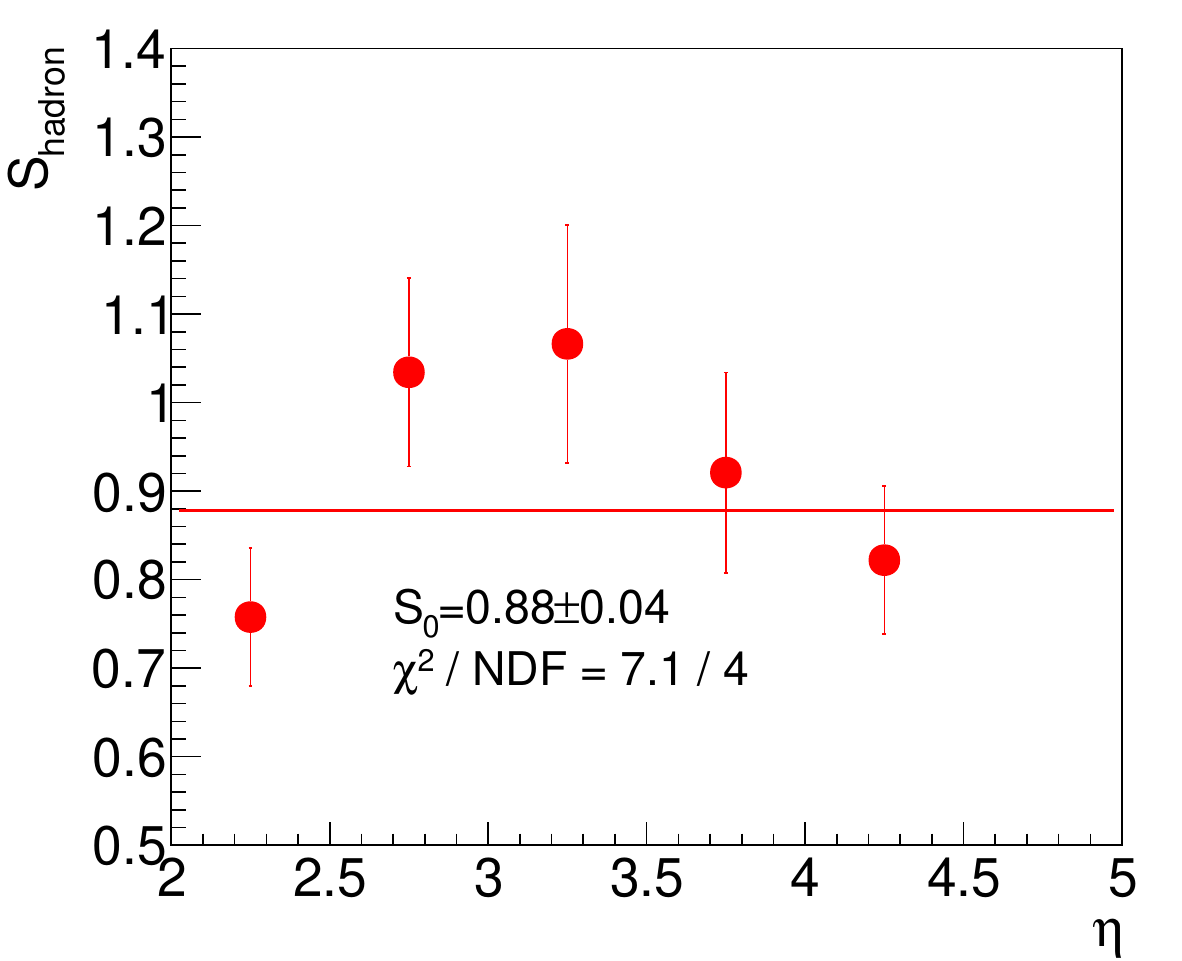}
}
\caption{An example fit of the multiplicity distributions (left) measured by LHCb \cite{LHCb:2014wmv} in $2<\eta<2.5$ range of rapidity. The fit excludes points $n<3$. The extracted entropies as functions of rapidity (right).}
\label{fig:LHCbfit}
\end{figure}

If the charged-particle multiplicity distributions are measured in narrow rapidity windows such that the $\mu \approx$ const. approximation holds, then Eq. (\ref{eq:exp_pn}) can be fitted to the data. A comparison with the LHCb measurement \cite{LHCb:2014wmv} was performed, and an example fit is shown in the left panel of Fig. \ref{fig:LHCbfit}. The extracted Shannon entropy is presented in the right panel of Fig. \ref{fig:LHCbfit}. (Comparison with data measured using expanding rapidity windows is also possible but requires further developments of the presented methods, as discussed in Sec. \ref{sec:rapwindow}.)

The distribution derived from the POME can describe the data, and the extracted entropy is approximately constant. The parameter values and the $\chi^2/$NDF are given in the plot. The POME-based description of the multiplicity distributions is statistically acceptable. The entropy is observed to be constant which is in agreement with the H1 result \cite{H1:2020zpd}.

\section{Towards large rapidity intervals}
\label{sec:rapwindow}

The above derivation assumed that the mean is constant which is only valid in a narrow rapidity interval. To describe distributions measured in wider rapidity intervals, two main approaches can be considered. One can compose multiple exponential distributions from Eq. (\ref{eq:exp_pn}) into a single distribution. Another way is to generalize the constraint so the mean depends on the variable as $\mu=\mu(n)$. In this paper, we follow the first approach, while the second is left for discussion in a future work.

Let us consider that within a narrow rapidity window, the exponential distribution can describe the measured multiplicity. To extend the rapidity range to twice its width, one could calculate the convolution of the distribution with itself. Repeating this process $N$ times that would ultimately leads to the Gamma distribution, denoted by $\Gamma(k=N,\lambda)$, where $\lambda=1/\mu$ is the rate parameter, and $k$ is so-called the shape parameter. The Gamma distribution is known to be the scaling function of the negative binomial distribution. Indeed, with the Poisson-Gamma mixture or Poisson transformation, which transforms a continuous distribution into a discrete one, it can be shown if $Y\sim\Gamma(k,\lambda)$ and $X\sim\textmd{Poisson}(\mu)$ as $\mu=Y$, then
\begin{equation}
    P(X=x|\mu) = \textmd{NBD}(k,\overline{n}=k/\lambda).
\end{equation}
Consequently, it is shown that a natural explanation for the widely observed negative binomial distribution (NBD) shape of charged particle multiplicities emerges from the POME-based derivation: the origin of the NBD shape lies in the convolution of maximal entropy distributions repeated $k=N$ times. This argument is more general than the well-known clan model \cite{Giovannini:1991ve}. Let us examine the relation between the parameters of the Gamma distribution and those of the NBD:
\begin{equation}
k_\Gamma = k_\textmd{NBD} \hspace{1.5cm} \lambda_\Gamma=\frac{k_\textmd{NBD}}{\overline{n}}.
\label{eq:gamma_nbd_relations}
\end{equation}
These relations have three consequences. First, $k$ should depend linearly on the width of the rapidity window; it is observed. Second, the entropy should saturate. It is expected that the entropy does not increase indefinitely as more phase space is covered -- this saturation behavior was already observed in the UA5 measurements \cite{SIMAK1988159,UA5:1988gup}.

Finally, $k$ should be a positive integer as it is the number of underlying or convoluted exponential distributions. However, this behavior is not confirmed by the experimental results. The reason could be short and long range correlations that would reduce entropy but were not considered in the derivation. Also, by convoluting exponential distributions, tacitly, a resolution is defined, which may not be valid across all regions. It is also important to note that the negative binomial parameters are strongly correlated, which may be a consequence of the relationship shown in Eq. (\ref{eq:gamma_nbd_relations}).

\section{Summary}

The derivation of the charged particle multiplicity distributions was presented based on the principle of maximal entropy. The derived distributions have a well-known functional form that describes the data well in a narrow rapidity ranges, where the constant-mean approximation can be considered to be valid. The results were compared to LHCb data and Shannon's entropy was extracted, found to be consistent with a constant value. This is in agreement with previous observations in deep inelastic $e^-p$ scattering by H1. The approach was then extended to wide rapidity ranges and was shown to explain the negative binomial shape of the measured distributions.

\section*{Acknowledgments}
S.L. is grateful for the support of the Miniatura 7 grant of the National Science Centre of Poland. S.L. thanks the hospitality and insightful discussions to Mateusz Plosko{\'n} and Jesus Guillermo Contreras Nuno. S.L. would like to thank the fruitful discussions to Krzysztof Kutak, Martin Hentschinski and Jacek Otwinowski.

\end{document}